\documentclass{osa-article}

\journal{osajournal}

\articletype{Research Article}

		\usepackage{textgreek}

		\usepackage{cuted}
		\catcode`@=11
		\def\nvphantom{\v@true\h@false\nph@nt}
		\def\nhphantom{\v@false\h@true\nph@nt}
		\MakeRobust{\vphantom}
		\def\nphantom{\v@true\h@true\nph@nt}
		\def\nph@nt{\ifmmode\def\next{\mathpalette\nmathph@nt}\else\let\next\nmakeph@nt\fi\next}
		\def\nmakeph@nt#1{\setbox\z@\hbox{#1}\nfinph@nt}
		\def\nmathph@nt#1#2{\setbox\z@\hbox{$\m@th#1{#2}$}\nfinph@nt}
		\def\nfinph@nt{\setbox\tw@\null \ifv@ \ht\tw@\ht\z@ \dp\tw@\dp\z@\fi\ifh@ \wd\tw@-\wd\z@\fi \box\tw@}
		\usepackage{stackengine}[2013-10-15]
		\usepackage{scalerel}
		
		\renewcommand{\sin}[1]{\text{sin}\!\left({#1}\right)}
		
		\renewcommand{\cos}[1]{\text{cos}\!\left({#1}\right)}

		\newcommand{\Real}[1]{\text{Re}\!\left({#1}\right)}

		\newcommand{\Imag}[1]{\text{Im}\!\left({#1}\right)}

		\newcommand{\diff}[1]{\text{d}{#1}}

		\newcommand{\integral}[4]{\int_{#1}^{#2}{#3}\ifthenelse{\isempty{#4}}{}{\,\text{d}{#4}}}
		\newcommand{\ointegral}[3]{\oint_{#1}{#2}\ifthenelse{\isempty{#3}}{}{\,\text{d}{#3}}}
		\usepackage{scalerel}[2016-12-29]

		\usepackage[makeroom]{cancel}
		
		\renewcommand{\leq}{\leqslant}
		\renewcommand{\geq}{\geqslant}
		\renewcommand{\;}{$,$}
		\renewcommand{\.}{\,\!}
		\newcommand*{\eq}[1]{\begin{align}#1\end{align}}
		
		\newcommand*{\weqb}[1]{\begin{strip}\rule{\dimexpr(0.5\textwidth-0.5\columnsep-0.4pt)}{0.4pt}\eq{#1}}
		\newcommand*{\weqe}[1]{\eq{#1}\par\hfill\rule[0.5\baselineskip]{\dimexpr(0.5\textwidth-0.5\columnsep-1pt)}{0.4pt}\end{strip}}

		\usepackage{etoolbox}
		\AtBeginEnvironment{eqnarray}{\setlength{\arraycolsep}{2pt}}

		\newcommand{\figwidth}{1\linewidth}
		
		\newcommand{\graphwidth}{1.2*\figwidth/sqrt(2)}
		\newcommand{\graphheight}{0.75*\graphwidth}
		\usepackage{pgfplots}
		\pgfplotsset{compat=1.9}
		\usepgfplotslibrary{fillbetween}
		\usepgfplotslibrary{patchplots}
		\usepgfplotslibrary{polar}
		\usetikzlibrary{patterns}
		\definecolor{blood}{rgb}{0.8,0,0}
		\definecolor{sodium}{rgb}{1,0.75,0}
		\definecolor{chlorophyll}{rgb}{0,0.8,0}
		\definecolor{cobalt}{rgb}{0.2,0.2,1}
		\definecolor{iodine}{rgb}{0.6,0.2,0.9}
		\pgfplotsset{colormap={Mathematicarainbow}{rgb=(0.431,0.137,0.5061);rgb=(0.259,0.184,0.702);rgb=(0.278,0.38,0.784);rgb=(0.365,0.561,0.737);rgb=(0.455,0.667,0.608);rgb=(0.561,0.725,0.471);rgb=(0.678,0.745,0.38);rgb=(0.792,0.718,0.333);rgb=(0.831,0.612,0.298);rgb=(0.835,0.42,0.239);rgb=(0.792,0.216,0.18)}}

		\usepackage{xifthen}
		\newcommand{\article}[6]{#1 (#2): \textit{#3}, #4 #5, \mbox{#6}}
		
		\newcommand{\articlesub}[4]{#1 (#2): \textit{#3}, submitted to #4}
		
		\newcommand{\book}[6]{#1 (#2): \textit{#3}, \ifthenelse{\isempty{#4}}{}{\ifthenelse{\equal{\detokenize{#4}}{\detokenize{2}}}{second}{\ifthenelse{\equal{\detokenize{#4}}{\detokenize{3}}}{third}{\ifthenelse{\equal{\detokenize{#4}}{\detokenize{4}}}{fourth}{\ifthenelse{\equal{\detokenize{#4}}{\detokenize{5}}}{fifth}{\ifthenelse{\equal{\detokenize{#4}}{\detokenize{6}}}{sixth}{\ifthenelse{\equal{\detokenize{#4}}{\detokenize{7}}}{seventh}{\ifthenelse{\equal{\detokenize{#4}}{\detokenize{8}}}{eighth}{\ifthenelse{\equal{\detokenize{#4}}{\detokenize{9}}}{ninth}{#4}}}}}}}} edition, }#5\ifthenelse{\isempty{#6}}{}{, \mbox{#6}}}
		\newcommand{\bookarticle}[7]{#1 (#2): \textit{#3}, in \textit{#4}\ifthenelse{\isempty{#5}}{}{ #5}, #6, \mbox{#7}}

		\newcommand{\Ec}{\vec{E}^\text{\,c}}
		\newcommand{\Eco}{\vec{E}^\text{\,c}_0}
		\newcommand{\Ee}{\vec{E}^\text{\,e}}
		
		\newcommand{\Ei}{\vec{E}^\text{\,i}}
		\newcommand{\Es}{\vec{E}^\text{\,s}}

		\newcommand{\FEs}{\vec{\mathcal{E}}^\text{\,s}}

		\DeclareRobustCommand{\hbar}{\mathrlap{\mspace{-1mu}\bar{\phantom{x}}}h}
		\newcommand{\ke}{k_\text{e}}
		\newcommand{\km}{k_\text{m}}
		\newcommand{\kx}{k_x}
		\newcommand{\ky}{k_y}
		\newcommand{\kz}{k_z}
		
		\renewcommand{\ne}{n_\text{e}}
		\newcommand{\nm}{n_\text{m}}
		\newcommand{\np}{n_\text{p}}
		\newcommand{\Qj}{\mathcal{Q}_j}
		\newcommand{\Ql}{\mathcal{Q}_\ell}
		\newcommand{\Qq}{\mathcal{Q}_q}
		\newcommand{\evalr}{(\vec{r})}
		\newcommand{\evalrj}{(\vec{r}_j)}
		
		\newcommand{\evalrl}{(\vec{r}_\ell)}

		\newcommand{\evalrjrl}{(\vec{r}_j,\vec{r}_\ell)}
		\newcommand{\evalrlrj}{(\vec{r}_\ell|\vec{r}_j)}

		\newcommand{\evalrrp}{(\vec{r},\vec{r}\,')}
		\newcommand{\evalrpp}{(\vec{r}\,'')}

		\newcommand{\Sfwd}{S(0)}
		\newcommand{\Stheta}{S(\theta)}
		\newcommand{\xj}{x_j}
		\newcommand{\xl}{x_\ell}
		\newcommand{\yj}{y_j}
		\newcommand{\yl}{y_\ell}
		\newcommand{\zj}{z_j}
		\newcommand{\zl}{z_\ell}

\begin{document}\sloppy

\twocolumn[{%
\begin{@twocolumnfalse}

\title{Multiple-scattering model for the effective refractive index of dense suspensions of forward-scattering particles}

\author{Alexander Nahmad-Rohen\authormark{1,*} and Augusto Garc\'ia-Valenzuela\authormark{1}}

\address{\authormark{1}Instituto de Ciencias Aplicadas y Tecnolog\'ia, Universidad Nacional Aut\'onoma de M\'exico, Apartado Postal 70186, Ciudad de M\'exico 04510, M\'exico}

\email{\authormark{*}alexander.nahmad@icat.unam.mx}

\begin{abstract}
We present a multiple-scattering model for the effective refractive index of an arbitrarily dense suspension of forward-scattering particles. The model provides a very simple formula for the effective refractive index of such a suspension and reproduces with high accuracy available experimental results. Furthermore, the derivation we present herein is mathematically transparent and enables us to obtain information on the underlying physical processes rather than obscuring them. We also provide insight into the extent of the model's validity and a simple way to determine whether or not it will be valid for an arbitrary suspension. Due to its simplicity, analytical closedness and wide range of applicability, we believe the model can be used as a diagnostic tool for complex materials of vastly different natures.
\end{abstract}

\end{@twocolumnfalse}}]

\section{Introduction}

The study of optically complex materials ---that is, heterogeneous mixtures of multiple components, each with its own optical properties, such as colloids--- is inherently interesting and has applications in a wide variety of fields, including atmospheric physics\cite{ref-vandeHulst-LSSP,ref-Irvine-JGR71}, astronomy\cite{ref-vandeHulst-LSSP,ref-Irvine-JGR71,ref-Lumme-I126}, communications\cite{ref-vandeHulst-PSPIE2052}, materials science\cite{ref-Pabst-JECS41,ref-Hribalova-JECS41}, biology\cite{ref-vandeHulst-PSPIE2052,ref-NahmadRohen-JOSAA38} and medicine\cite{ref-vandeHulst-PSPIE2052,ref-NahmadRohen-JOSAA38,ref-NahmadRohen-PS91}. However, it is also, by its very nature, complicated. Effective-medium theory provides one way to study them; it involves the calculation of an effective electromagnetic response (including an effective refractive index) which encodes the optical and geometric properties of the material's components. Within effective-medium theory, multiple models have been proposed throughout the years, starting with the Maxwell-Garnett model\cite{ref-Garnett-PTRSA203}, which neglects scattering by the different components, and culminating in van de Hulst's well-known formula\cite{ref-vandeHulst-LSSP}, which is consistent to first order in the number density of the scattering components of the material and is a benchmark against which other models are very frequently compared. Most of these models feature the scattering amplitude of the particles that make up a complex material, typically in the forward direction, because they are fundamentally scattering models. Therefore, to properly study complex materials it is strictly necessary to understand how their various components scatter light.

It is important to emphasise the limitations of effective-medium models. For a material consisting of particles of the order of the wavelength or larger, they are appropriate for describing the refraction of light at the material's edge and the propagation of light through the material (with an effective propagation constant, from which the aforementioned effective refractive index can be obtained), but not reflection at the material's edge\cite{ref-Bohren-JAS43,ref-GutierrezReyes-JPCB118}. This limitation is due to the inherent non-locality of such systems\cite{ref-Barrera-PRB75}.

Nevertheless, within its limits of applicability, the effective refractive index is a useful tool to optically characterise complex materials, which we shall here take to consist of a number of particles suspended in a homogeneous medium called the matrix. It depends on predictable and often measurable ways on the amount of particles per unit volume (or, equivalently, the volume fraction, which is the volumetric concentration of particles or, in other words, the fraction of the suspension's volume taken up by the particles; this quantity is often much more useful than the particle number density), the refractive indices of the particles and the matrix, the size of the particles and the wavelength of light.

The applicability of existing models to dense suspensions (volume fractions higher than about 0.1) has been widely debated\cite{ref-Hribalova-JECS41,ref-Ishimaru-JOSA72,ref-Tsang-JQSRT224}. The principal reason for this is the existence of dependent scattering --- that is, the field scattered by any one particle in the suspension depends on the fields scattered by all other particles in the suspension, leading to a highly coupled system which is not always easy to treat mathematically. ``Dependent scattering'' is presently an ambiguous term for which no precise definition exists, as different definitions have been given by different authors\cite{ref-Hespel-JOSAA18,ref-Durant-JOSAA24,ref-Mishchenko-OSAC1}; here we will take the side of those who define it as a non-linear dependence of the effective refractive index on the particle number density\cite{ref-Hespel-JOSAA18,ref-Durant-JOSAA24,ref-Ishimaru-JOSA72}.

Despite van~de~Hulst's formula having been shown to be widely applicable and sometimes even be accurate for dense suspensions of particles\cite{ref-Alexander-JCSFT277,ref-Meeten-OC134,ref-Meeten-MST8}, its validity for such suspensions remained controversial until recently\cite{ref-NahmadRohen-JOSAA38}, when it was rigorously shown that for suspensions of large tenuous particles it can be valid up to volume fractions of about 0.45 with relatively small errors. While said proof will hopefully end such controversy, it was too limited by making strong assumptions about the particle and matrix properties. Furthermore, while the mathematics of said proof was unambiguous, the physics of light-matter interaction in dense suspensions was somewhat unclear.

A very recently proposed model\cite{ref-GarciaValenzuela-JQSRT} taking into account dependent scattering is applicable to dense suspensions of arbitrary particles. Once again, however, the complicated mathematics involved in the derivation of the relatively simple formula obscures the physical processes involved. Furthermore, while the model is applicable to a very wide range of possible suspensions of randomly positioned particles, the formula is computationally costly. Other works are even more mathematically complicated and thus hide the underlying physics further\cite{ref-Tsang-RS35}, are applicable only to thin films rather than bulk matter\cite{ref-Pecharroman-JAP93} or require extremely tenuous particles\cite{ref-Looyenga-P31,ref-Twersky-JOSA60}.

Thus, most effective-medium models remain unsatisfactory for the description of the optical properties of dense suspensions; either they are restricted to small particles and/or dilute suspensions, they involve complicated mathematical expressions which hide the underlying physics and cannot be solved analytically, or both. Thus, an analytical formula for the effective refractive index of a dense suspension remains elusive.

In this work, we present a detailed derivation of a formula for the effective refractive index of suspensions of particles which scatter light mainly in the forward direction, which we call ``highly-forward-scattering particles'' here. The formula takes into account dependent scattering and will be seen to be valid for surprisingly dense suspensions (up to, and possibly exceeding, volume fractions of 0.4 in some cases; note that 0.4 is already close to the maximum possible density for identical spherical particles, which is about 0.74). Our model is based, initially, on the quasi-crystalline approximation\cite{ref-Tsang-SEW}. In the process, we provide insight into the theory of the effective refractive index in general, into when dependent scattering arises and how it affects our treatment of the problem, and into the physics of the interaction of highly-forward-scattering particles with light. For example, it shall be seen that particle correlation functions cease to become important as the particles become increasingly forward-scattering. Our final formula is essentially as simple as van~de~Hulst's, and we have aimed to make both the mathematics and the physics as transparent as possible. This is in contrast to the traditional approach to the quasi-crystalline approximation, which is to obtain numerical solutions to a self-consistent integral equation, which is computationally costly and provides no physical insight into the relationship between the effective refractive index and the parameters of the suspension.

We then show that our model reproduces experimental results available in the scientific literature for suspensions of forward-scattering particles with very high accuracy even at high volume fractions. We further discuss when our formula is valid and when it is not, and we provide examples of real systems to which it is applicable, contributing to optical diagnosis of such systems. An in-depth analysis of the application of our model to some such systems, however, remains a topic for future articles.

\onecolumn

\section{General multiple-scattering theory}\label{sec-E}

Let us consider a plane wave \smash{$\Ei=\Ei_0\,e^{i\km z}$} travelling in the $z$ direction through a homogeneous medium (which we call the matrix) with refractive index $\nm$ and entering at \smash{$z=0$} a region consisting of a suspension of $N$ identical homogeneous particles suspended in the matrix. Here \smash{$\km=\nm k$}, where $k$ is the wave number in vacuum.

The total field at an arbitrary point $\vec{r}$ in the matrix is
\eq{\vec{E}\evalr=\Ei\evalr+\Es\evalr,\label{eq-generalE}}
where \smash{$\Es$} is the total scattered field, equal to the sum of the fields scattered by the individual particles of the suspension.

The field scattered by any given particle, which we identify with the index $j$ and whose centre is located at \smash{$\vec{r}_j=(\xj,\yj,\zj)$}, is a function of the field \smash{$\Ee_j$} that excites that particle. We may write
\eq{\Es\evalr=\sum_{j=1}^N\Qj[\Ee_j](\vec{r}),\label{eq-generalEsj}}
where $\Qj$ is a linear operator which tells us how the $j$th particle scatters light. $\Qj$ is, strictly speaking, an integral operator where the integration occurs over the whole volume available to the particle (i.e.~the whole volume of the suspension of particles)\cite{ref-Tsang-SEW,ref-Mishchenko-PR632}:
\eq{\Qj[\Ee_j](\vec{r})=\displaystyle\integral{}{}{\!\!\!\integral{}{}{\!\!\!\integral{}{}{\!\!\!\integral{}{}{\!\!\!\integral{}{}{\!\!\!\integral{}{}{\mathcal{G}\evalrrp\mathcal{T}_j(\vec{r}\,'-\vec{r}_j,\vec{r}\,''-\vec{r}_j)\Ee_j\evalrpp}{x''}}{y''}}{z''}}{x'}}{y'}}{z'},}
where , $\mathcal{T}_j$ is the $j$th particle's dyadic transition operator and $\mathcal{G}$ is the dyadic Green's function (throughout this work, an integral without integration limits is understood to be over the entirety of the suspension's volume -- or, in the case of Fourier integrals, from $-\infty$ to $\infty$). $\mathcal{T}_j$ is such that the integrand is zero outside the particle's volume.

In turn, within the $j$th particle's volume, \smash{$\Ee_j$} is given by
\eq{\Ee_j\evalr=\Ei\evalr+\sum_{\ell\neq j}\Ql[\Ee_\ell]\evalr,\label{eq-generalEej}}
where \smash{$\Ee_\ell$} is the field exciting the $\ell$th particle within the $\ell$th particle's volume, yielding a strongly coupled system of equations for the $N$ exciting fields. Combining equations~\ref{eq-generalE}, \ref{eq-generalEsj} and~\ref{eq-generalEej}, we obtain an expression for the total field in the matrix:
\eq{\vec{E}\evalr=\Ei\evalr+\sum_{j=1}^N\Qj[\Ei]\evalr+\sum_{j=1}^N\sum_{\ell\neq j}\Qj\!\left[\Ql[\Ee_\ell]\right]\!\evalr.}
We could continue iterating on the above, writing
\eq{\Ee_\ell\evalr=\Ei\evalr+\sum_{q\neq\ell}\Qq[\Ee_q]\evalr}
and so on, yielding
\eq{\vec{E}\evalr=\Ei\evalr+\sum_{j=1}^N\Qj[\Ei]\evalr+\sum_{j=1}^N\sum_{\ell\neq j}\Qj\!\left[\Ql[\Ei]\right]\!\evalr+\sum_{j=1}^N\sum_{\ell\neq j}\sum_{q\neq\ell}\Qj\!\left[\Ql\!\left[\Qq[\Ei]\right]\vphantom{\frac{b}{b}}\!\right]\!\evalr+\ldots.\label{eq-Egeneral}}

The beauty of this representation of the total field is that each scattering order is presented as a separate term; beginning with the terms with sums in them, the first term is the incidentfield scattered once (by the $j$th particle), the second term is the incident field scattered twice (first by the $\ell$th particle and then by the $j$th particle) and so on. Because the term corresponding to $s$th-order scattering (for some \smash{$s\in\mathbb{N}$}) has $s$ nested sums, it is proportional to $N(N-1)^{s-1}$, which, for large $N$, is approximately equal to $N^s$. This is because the outermost sum in each term (with index $j$) is over the entire collection of particles and every other sum (with indices other than $j$) is over all particles except one.

Since equation~\ref{eq-Egeneral} is an infinite series, we truncate it at some value of $s$ (this is usually done at the first-order-scattering term). All terms on the right side of the truncated series are then dependent on \smash{$\Ei$} except for the last one, which is dependent on the field exciting the particle with the index corresponding to the final ($s$th) nested sum in that term. No approximations have been made thus far, but calculating the aforementioned exciting field in the final term of the truncated series can sometimes be difficult.

The usual approach at this stage is to take the configurational average of \smash{$\vec{E}$} --- that is, the average of \smash{$\vec{E}$} over all possible configurations (positions and orientations) of the particles. This is appropriate because we do not know the exact positions and orientations of the particles. This average is the coherent component of the light transmitted through the suspension and will be called the coherent field throughout this work; we denote it by \smash{$\Ec\equiv\langle\vec{E}\rangle$}.

The averages are calculated by integrating over the entire volume of the suspension (we neglect writing the integrals over particle orientations for the sake of clarity):
\eq{\Ec\evalr&=\Ei\evalr+\sum_{j=1}^N\integral{}{}{\!\!\!\integral{}{}{\!\!\!\integral{}{}{g\evalrj\Qj[\Ei]\evalr}{\xj}}{\yj}}{\zj}\nonumber\\
&\hphantom{=\,\,}+\sum_{j=1}^N\sum_{\ell\neq j}\integral{}{}{\!\!\!\integral{}{}{\!\!\!\integral{}{}{g\evalrj\Qj\!\left[\integral{}{}{\!\!\!\integral{}{}{\!\!\!\integral{}{}{g\evalrlrj\Ql[\Ei]\evalr}{\xl}}{\yl}}{\zl}\right]\!\evalr}{\xj}}{\yj}}{\zj}+\ldots.\label{eq-Ecint}}
In the above, $g\evalrj$ is the probability density that the $j$th particle is located within a volume differential $\diff{\xj}\,\diff{\yj}\,\diff{\zj}$ around $\vec{r}_j$. \smash{$g\evalrlrj=c\evalrjrl g\evalrl$} is the conditional probability density that the $\ell$th particle is located within a volume differential $\diff{\xl}\,\diff{\yl}\,\diff{\zl}$ around $\vec{r}_\ell$ given that the $j$th particle is located at $\vec{r}_j$; $c\evalrjrl$ is the two-particle correlation function. Subsequent terms will include $s$-particle correlation functions with \smash{$s>2$}.

It is here implicit that, when we write $g\evalrlrj$, integration over all particles other than the $j$th and $\ell$th ones has already been performed; this is not shown for clarity. The analogous is true for higher-order terms.

The calculation thus becomes increasingly complicated and impractical the more terms of the series are taken. It is usual to truncate the series (as stated, at the first-order term) and replace the \smash{$\Ei$} in the last remaining term with \smash{$\Ec$}, which is called the effective-field approximation. This yields a model which is self-consistent to first order in the particle volume density. However, this is not the only approach.

If all particle orientations occur with equal probability and the particles are homogeneously distributed over the volume $V$ of the suspension, then, by rotational and translational symmetry, \smash{$\Ec$} must be a plane wave propagating in the $z$ direction, just like \smash{$\Ei$}, but with an effective propagation constant \smash{$\ke=\ne k$}, where $\ne$ is the effective refractive index of the suspension and is what we are after. We write \smash{$\Ec\evalr=\Eco e^{i\ke z}$}.

\section{The model}\label{sec-model}

Here we will truncate the series at the 2nd-order-scattering term and focus on the average field exciting the $j$th particle. This is called the quasi-crystalline approximation\cite{ref-Tsang-SEW}. This will turn equation~\ref{eq-generalEej} into an expression correct to first order in the number density $\rho$ of the particles (or, equivalently, in the volume fraction $f$), which we can then insert into equation~\ref{eq-generalEsj} to yield an expression for \smash{$\Ec$} correct to second order in $\rho$. We will not explicitly calculate this expression for the coherent field, since we are only interested in finding $\ne$, but this shows that our model is self-consistent to second order in $\rho$.

Once our self-consistent expression for the average field exciting the $j$th particle is established, we will solve it for $\ke$ in order to obtain a simple formula for the effective refractive index $\ne$ of the suspension. We will first assume the particles are spherical and then generalise the model to suspensions of arbitrarily shaped particles.

\subsection{The average field exciting the $j$th particle}\label{subsec-Eej}

The term on which $\Qj$ operates is the average field exciting the $j$th particle, \smash{$\langle\Ee_j\rangle_{{\textstyle\mathstrut}j}$} (the subindex $j$ outside the average brackets indicates that this configurational average is over all particles other than the $j$th one; the $j$th particle's position remains fixed). This is given by taking the configurational average on both sides of equation~\ref{eq-generalEej}:
\eq{\langle\Ee_j\rangle_{{\textstyle\mathstrut}j}\evalr=\Ei\evalr+\sum_{\ell\neq j}\integral{}{}{\!\!\!\integral{}{}{\!\!\!\integral{}{}{g\evalrlrj\Ql[\Ee_\ell]\evalr}{\xl}}{\yl}}{\zl}.}

Instead of iterating as in the preceding section in order to find an expression for \smash{$\langle\Ee_j\rangle_{{\textstyle\mathstrut}j}$} in powers of $\rho$, we will henceforth assume that the field exciting the $\ell$th particle can be replaced by its configurational average (fixing the $\ell$th particle's position), which, by rotational and translational symmetry, is a plane wave, just like the coherent field:
\eq{\langle\Ee_j\rangle_{{\textstyle\mathstrut}j}\evalr=\Ei\evalr+\sum_{\ell\neq j}\integral{}{}{\!\!\!\integral{}{}{\!\!\!\integral{}{}{g\evalrlrj\Ql[\langle\Ee_\ell\rangle_{{\textstyle\mathstrut}\ell}]\evalr}{\xl}}{\yl}}{\zl}.\label{eq-EejEsl}}
This is the self-consistency equation in the quasi-crystalline approximation\cite{ref-Tsang-SEW}. The field scattered by the $\ell$th particle, denoted by \smash{$\Es_\ell$}, will thus, from now on, be assumed to be the field scattered by a particle on which the plane wave \smash{$\langle\Ee_\ell\rangle_{{\textstyle\mathstrut}\ell}$} is incident:
\eq{\Es_\ell\evalr=\Ql[\langle\Ee_\ell\rangle_{{\textstyle\mathstrut}\ell}]\evalr.}

If the particles scatter mostly in the forward direction, then the field scattered by the $\ell$th particle only contributes to the field at the $j$th particle's position, \smash{$\vec{r}_j=(\xj,\yj,\zj)$}, if \smash{$\zl+a\leq\zj-a$} (where $a$ is a measure of the size of a particle measured from the particle's centroid; for example, for spherical particles $a$ is their radius), so we insert a Heaviside function \smash{$\Theta(\zj-\zl-2a)$} in the integral in equation~\ref{eq-EejEsl}, which is equivalent to changing the upper integration limit in the integral over $\zl$:
\eq{\langle\Ee_j\rangle_{{\textstyle\mathstrut}j}\evalr=\Ei\evalr+\sum_{\ell\neq j}\integral{0}{\zj-2a}{\integral{}{}{\!\!\!\integral{}{}{g\evalrlrj\Es_\ell\evalr}{\xl}}{\yl}}{\zl}.}
By the same logic, we write \smash{$g\evalrlrj\approx1/V$}. Since, a priori, there is no constraint on the position of any given particle (ignoring the presence of all other particles), we have \smash{$g\evalrl=1/V$}. Thus, we have the approximation \smash{$c\evalrjrl\approx1$} in equation~\ref{eq-EejEsl}. We justify this form of $g\evalrlrj$ as follows: In the simplest case, $c\evalrjrl$ would be the hole correlation function, where the probability that the $\ell$th particle is at $\vec{r}_\ell$ is zero in a vicinity of the $j$th particle (because particles cannot overlap) and uniform elsewhere; the simplification of this to \smash{$c\evalrjrl=1$} occurs because we are only considering particles centred at \smash{$\zl\leq\zj-2a$}, where there is no restriction because the $j$th particle is absent in that region. Even if \smash{$c\evalrjrl$} were some other function, such as the Percus-Yevick correlation function\cite{ref-Ishimaru-JOSA72}, this would only introduce a significant correction in a small (of the order of the particle size) region close to $\vec{r}_j$, which is negligible compared to the size of the space \smash{$\zl+a\leq\zj-a$}. We therefore have
\eq{\langle\Ee_j\rangle_{{\textstyle\mathstrut}j}\evalr=\Ei\evalr+\sum_{\ell\neq j}\frac{1}{V}\integral{0}{\zj-2a}{\integral{}{}{\!\!\!\integral{}{}{\Es_\ell\evalr}{\xl}}{\yl}}{\zl}.\label{eq-Eej}}

\subsection{The average field scattered by the $\ell$th particle}\label{subsec-Esl}

We now turn our attention to \smash{$\Es_\ell$}. At the \smash{$z=\zl+a$} plane, called the expansion plane of the $\ell$th particle, the two-dimensional Fourier transform of the field scattered by the $\ell$th particle, \smash{$\Es_\ell$}, is
\eq{\FEs_\ell(\kx,\ky,\zl+a)=\frac{1}{2\pi}\integral{}{}{\!\!\!\integral{}{}{\Es_\ell(x,y,\zl+a)\,e^{-i(\kx x+\ky y)}}{x}}{y},\label{eq-FT}}
where we have taken the convention (usual in mathematics) that the direct and inverse Fourier transforms both have a factor $1/\sqrt{2\pi}$ per variable. The inverse Fourier transform of \smash{$\FEs_\ell$} at the expansion plane is, of course,
\eq{\Es_\ell(x,y,\zl+a)=\frac{1}{2\pi}\integral{}{}{\!\!\!\integral{}{}{\FEs_\ell(\kx,\ky,\zl+a)\,e^{i(\kx x+\ky y)}}{\kx}}{\ky}.\label{eq-IFT}}

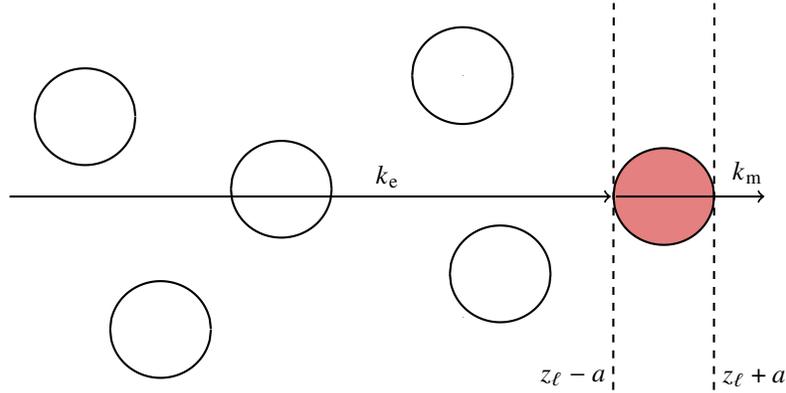
\begin{figure}[t!]
	\begin{center}{
		\centering
		\begin{tikzpicture}
			\begin{axis}[
				width=0.75*\figwidth,
				height=0.45*\figwidth,
				axis line style=thick,
				legend cell align=left,
				legend style={draw=none},
				xmin=-1.9,
				xmax=1.3,
				xticklabels=,
				axis x line=middle,
				x axis line style={draw=none},
				x label style={at={(current axis.right of origin)},anchor=west},
				xlabel=,
				ymin=-0.8,
				ymax=1,
				yticklabels=,
				scaled ticks=false,
				major tick length=0,
				axis y line=center,
				y axis line style={draw=none},
				y label style={at={(current axis.above origin)},anchor=south},
				ylabel=,
				disabledatascaling
			]
				\addplot[smooth,thick,black,domain=0.6:1,samples=101,name path=pjp]{sqrt(0.2^2-(x-0.8)^2)};
				\addplot[smooth,thick,black,domain=0.6:1,samples=101,name path=pjm]{-sqrt(0.2^2-(x-0.8)^2)};
				\addplot[blood,opacity=0.5]fill between[of=pjm and pjp];
				\addplot[smooth,thick,black,domain=-0.05:0.35,samples=101]{sqrt(0.2^2-(x-0.15)^2)-0.32};
				\addplot[smooth,thick,black,domain=-0.05:0.35,samples=101]{-sqrt(0.2^2-(x-0.15)^2)-0.32};
				\addplot[smooth,thick,black,domain=-0.2:0.2,samples=101]{sqrt(0.2^2-x^2)+0.5};
				\addplot[smooth,thick,black,domain=-0.2:0.2,samples=101]{-sqrt(0.2^2-x^2)+0.5};
				\addplot[smooth,thick,black,domain=-0.92:-0.52,samples=101]{sqrt(0.2^2-(x+0.72)^2)+0.03};
				\addplot[smooth,thick,black,domain=-0.92:-0.52,samples=101]{-sqrt(0.2^2-(x+0.72)^2)+0.03};
				\addplot[smooth,thick,black,domain=-1.4:-1,samples=101]{sqrt(0.2^2-(x+1.2)^2)-0.55};
				\addplot[smooth,thick,black,domain=-1.4:-1,samples=101]{-sqrt(0.2^2-(x+1.2)^2)-0.55};
				\addplot[smooth,thick,black,domain=-1.7:-1.3,samples=101]{sqrt(0.2^2-(x+1.5)^2)+0.33};
				\addplot[smooth,thick,black,domain=-1.7:-1.3,samples=101]{-sqrt(0.2^2-(x+1.5)^2)+0.33};
				\addplot[thick,dashed,black,domain=0.599:0.601,samples=2]{800*x-480};
				\draw[draw=none]node at(axis cs:0.44,-0.75){\textcolor{black}{$\zl-a$}};
				\addplot[thick,dashed,black,domain=0.999:1.001,samples=2]{800*x-800};
				\draw[draw=none]node at(axis cs:1.16,-0.75){\textcolor{black}{$\zl+a$}};
				\addplot[smooth,thick,->,black,domain=-1.8:0.59,samples=2]{0};
				\draw[draw=none]node at(axis cs:-0.3,0.08){\textcolor{black}{$\ke$}};
				\addplot[smooth,thick,->,black,domain=0.61:1.2,samples=2]{0};
				\draw[draw=none]node at(axis cs:1.13,0.1){\textcolor{black}{$\km$}};
			\end{axis}
		\end{tikzpicture}
		\captionsetup{singlelinecheck=off}
		\caption[.]{The average exciting field (black arrow) travels through the effective medium up to the plane \smash{$\zl-a$} (i.e.~just before the $\ell$th particle, shown here in red), but locally the field is immersed in the matrix, so the exciting field travels through the matrix once it reaches this plane.}
		\label{fig-localmedium}
	}
	\end{center}
\end{figure}

As mentioned earlier, we will assume any given particle is excited by the average exciting field, which, as stated at the beginning of section~\ref{sec-model}, is a plane wave travelling in the $z$ direction; we propose
\eq{\langle\Ee_\ell\rangle_{{\textstyle\mathstrut}\ell}(x,y,z\geq\zl-a)=\Ee_{\ell,0}\,e^{i\left(\ke(\zl-a)+\km(z-\zl+a)\vphantom{b^2}\right)}\label{eq-Eelphase}}
for some amplitude \smash{$\Ee_{\ell,0}$}. The reasoning behind the strange-looking phase is as follows (see figure~\ref{fig-localmedium}): up to the plane \smash{$\zl-a$} (i.e.~the plane immediately before the particle), the exciting field may be seen as travelling through the effective medium (with an effective propagation constant $\ke$), as it is the result of scattering by all the particles located before this plane, which are randomly positioned and oriented; locally, however, the $\ell$th particle is embedded in the matrix, so we must take the matrix propagation constant $\km$ starting from \smash{$\zl-a$}. Since the expansion plane is located after the \smash{$z=\zl-a$}, this applies to the Fourier transforms above.

The field scattered by the $\ell$th particle inherits the phase of the field exciting it. Therefore, at the expansion plane we have
\eq{\Es_\ell(x,y,\zl+a)=\Es_{\ell,0}(x,y,\zl+a)\,e^{i\left(\ke(\zl-a)+2\km a\vphantom{b^2}\right)}}
for some field \smash{$\Es_{\ell,0}$}. Therefore,
\eq{\FEs_\ell(\kx,\ky,\zl+a)=\frac{e^{i\left(\ke(\zl-a)+2\km a\vphantom{b^2}\right)}}{2\pi}\integral{}{}{\!\!\!\integral{}{}{\Es_{\ell,0}(x,y,\zl+a)\,e^{-i(\kx x+\ky y)}}{x}}{y}.}

We can write \smash{$\FEs_\ell$} in terms of the field scattered by a hypothetical particle centred at $(0,0,-a)$ with the expansion plane passing through the origin of coordinates, which will make further calculations simpler. To do so, we employ the coordinate change \smash{$\xi:=x-\xl$}, \smash{$\eta:=y-\yl$}, \smash{$\zeta:=z-(\zl+a)$}. This yields
\eq{\FEs_\ell(\kx,\ky,0)=\frac{e^{i\left(-\kx\xl-\ky\yl+\ke(\zl-a)+2\km a\vphantom{b^2}\right)}}{2\pi}\integral{}{}{\!\!\!\integral{}{}{\Es_{\ell,0}(x,y,\zl+a)\,e^{-i(\kx\xi+\ky\eta)}}{\xi}}{\eta}.}
By defining
\eq{\FEs_{\ell,0}(\kx,\ky):=\frac{1}{2\pi}\integral{}{}{\!\!\!\integral{}{}{\Es_{\ell,0}(x,y,\zl+a)\,e^{-i(\kx\xi+\ky\eta)}}{\xi}}{\eta}}
(i.e.~\smash{$\FEs_{\ell,0}$} is the Fourier transform of \smash{$\Es_{\ell,0}$}), we obtain
\eq{\FEs_\ell(\kx,\ky,0)=\FEs_{\ell,0}(\kx,\ky)\,e^{i\left(-\kx\xl-\ky\yl+\ke(\zl-a)+2\km a\vphantom{b^2}\right)}.\label{eq-FEslFESl0}}

Because \smash{$\FEs_\ell(\kx,\ky,\zeta)$} must satisfy the Helmholtz equation and its source is at \smash{$\zeta\leq0$}, it must be a wave propagating in the $+\zeta$ direction\cite{ref-Collin-ARP}. Therefore, in order to find the scattered field beyond the expansion plane we simply write\footnote{In a previous work\cite{ref-NahmadRohen-JOSAA38}, we had a similar equation (numbered~10 in that work) where there was a typographical error; there (and in equations~11 and~15 of the same work), we wrote $\km$ instead of $\kz$. This had no consequences for the final calculations, but it might have misled readers.}
\eq{\FEs_\ell(\kx,\ky,\zeta>0)=\FEs_\ell(\kx,\ky,\zeta=0)\,e^{i\kz\zeta},\label{eq-pwe}}
where \smash{$\kz=\sqrt{\km\.^2-\kx\.^2-\ky\.^2}$}. Using equations~\ref{eq-FEslFESl0} and~\ref{eq-pwe},
\eq{\Es_\ell(x,y,z>\zl+a)=\frac{e^{i\left(\ke(\zl-a)+2\km a\vphantom{b^2}\right)}}{2\pi}\integral{}{}{\!\!\!\integral{}{}{\FEs_{\ell,0}(\kx,\ky)\,e^{i\left(\kx(x-\xl)+\ky(y-\yl)+\kz(z-\zl-a)\vphantom{b^2}\right)}}{\kx}}{\ky}.}
This is a plane-wave expansion of the scattered field, since the integrand is a collection of plane waves travelling in directions given by the values of $\kx$ and $\ky$.

Taking the conditional configurational average (with the $j$th particle's position fixed) on both sides (which, as we know, equates to inserting the Heaviside function on the right and integrating with respect to the position of the $\ell$th particle, with the integral being weighted by the probability density of the position, which we have seen is approximately equal to $1/V$), we obtain
\eq{\langle\Es_\ell\rangle_{{\textstyle\mathstrut}j}\evalr&=\frac{e^{2i\km a}}{2\pi V}\int\!\!\!\int\!\!\!\integral{0}{\zj-2a}{e^{i\left((\ke-\kz)\zl-\ke a\right)}}{\zl}\integral{}{}{e^{-i\kx\xl}}{\xl}\integral{}{}{e^{-i\ky\yl}}{\yl}\,\times\nonumber\\
&\hphantom{=\,\,\frac{e^{2i\km a}}{2\pi V}\int\!\!\!\int\!\!\!}\times\FEs_{\ell,0}(\kx,\ky)\,e^{i\left(\kx x+\ky y+\kz(z-a)\vphantom{b^2}\right)}\,\diff{\kx}\,\diff{\ky}.}
But
\eq{\integral{}{}{e^{-i\kx\xl}}{\xl}=2\pi\delta(\kx)}
and similarly for the integral over $\yl$, so
\eq{\langle\Es_\ell\rangle_{{\textstyle\mathstrut}j}\evalr&=\frac{2\pi}{V}\,e^{i\km z}\integral{0}{\zj-2a}{e^{i(\ke-\km)(\zl-a)}}{\zl}\,\FEs_{\ell,0}(0,0)\nonumber\\
&=\frac{2\pi\left(e^{i\left(\km z+(\ke-\km)(\zj-3a)\vphantom{b^2}\right)}-e^{i\left(\km z-(\ke-\km)a\vphantom{b^2}\right)}\right)}{i(\ke-\km)V}\,\FEs_{\ell,0}(0,0),\label{eq-Eslavg}}
where for \smash{$\kx=\ky=0$} we have \smash{$\kz=\km$}.

In order to find \smash{$\FEs_{\ell,0}(0,0)$}, let us briefly return to equation~\ref{eq-IFT} and evaluate it in the far field of the particle --- that is, at some point $\vec{r}$ satisfying \smash{$|\vec{r}-\vec{r}_\ell|\gg\lambda,a$}, where $\lambda$ is the wavelength of the incident field. We stress that all calculations up to equation~\ref{eq-Eslavg} are valid for arbitrary $\vec{r}$; we take the far-field case only in the following few equations, and only to show that there is a simple relationship between \smash{$\FEs_{\ell,0}(0,0)$} and $S_\ell(0)$. Using the method of stationary phase\cite{ref-Collin-ARP,ref-Bohren-ASLSP},
\eq{\Es_\ell\evalr\approx-\frac{i\km\,\cos{\theta}e^{i\km|\vec{r}-\vec{r}_\ell|}}{|\vec{r}-\vec{r}_\ell|}\,\FEs_{\ell,0}\!\left(\km\,\sin{\theta}\cos{\varphi},\km\,\sin{\theta}\sin{\varphi}\vphantom{b^2}\!\right)\!,}
where $\theta$ is the polar angle of scattering (i.e.~the angle with respect to the $z$ direction) and $\varphi$ is the azimuthal angle. In the forward direction (\smash{$\theta=0$}), we have \smash{$\cos{\theta}=1$}, \smash{$\sin{\theta}=0$} and \smash{$|\vec{r}-\vec{r}_\ell|=z-\zl$}, whereby
\eq{\Es_\ell(z)=-\frac{i\km\,e^{i\km(z-\zl)}}{z-\zl}\,\FEs_{\ell,0}(0,0).}
Note the similarity between this expression and the scattering-amplitude-matrix expression for a particle's scattered field in the far-field regime:
\eq{\Es_\ell\evalr=\frac{e^{i\km|\vec{r}-\vec{r}_\ell|}}{-i\km|\vec{r}-\vec{r}_\ell|}\,S_\ell(\theta,\varphi)\Ee_{\ell,0}\,,}
where $S_\ell(\theta,\varphi)$ is the scattering-amplitude matrix of the particle and \smash{$\Ee_{\ell,0}$} is the amplitude of the field exciting the particle. In the forward direction, of course,
\eq{\Es_\ell(z)=\frac{e^{i\km(z-\zl)}}{-i\km(z-\zl)}\,S_\ell(0)\Ee_{\ell,0}\,,}
Comparing the two expressions, we arrive at the identity
\eq{\FEs_{\ell,0}(0,0)=-\frac{1}{\km\.^2}\,S_\ell(0)\Ee_{\ell,0}.\label{eq-FEsS0}}

Again we emphasise that we have briefly set our observation point in the far field of the $\ell$th particle only to make it possible to obtain the identity above. The remainder of this work does not require that the particles in the suspension be far from one another.

Using equations~\ref{eq-Eslavg} and~\ref{eq-FEsS0}, we have
\eq{\langle\Es_\ell\rangle_{{\textstyle\mathstrut}j}\evalr=\frac{2\pi i\left(e^{i\left(\km z+(\ke-\km)(\zj-3a)\vphantom{b^2}\right)}-e^{i\left(\km z-(\ke-\km)a\vphantom{b^2}\right)}\right)}{(\ke-\km)\km\.^2V}\,S_\ell(0)\Ee_{\ell,0}.\label{eq-avgEsl}}

\subsection{The effective refractive index}\label{subsec-ke}

Substituting equation~\ref{eq-avgEsl} into equation~\ref{eq-Eej},
\eq{\langle\Ee_j\rangle_{{\textstyle\mathstrut}j}\evalr=\Ei\evalr+\sum_{\ell\neq j}\frac{2\pi i\left(e^{i\left(\km z+(\ke-\km)(\zj-3a)\vphantom{b^2}\right)}-e^{i\left(\km z-(\ke-\km)a\vphantom{b^2}\right)}\right)}{(\ke-\km)\km\.^2V}\,S_\ell(0)\Ee_{\ell,0}.}

We now note that the average exciting fields have the same amplitude for all particles, since the particles are identical. Thus, we simply write \smash{$\Ee_0$} in place of \smash{$\Ee_{j,0}$} and \smash{$\Ee_{\ell,0}$}. In analogy to our expression for the average field exciting the $\ell$th particle (see equation~\ref{eq-Eelphase}), the average field exciting the $j$th particle is
\eq{\langle\Ee_j\rangle_{{\textstyle\mathstrut}j}\evalr=\Ee_0\,e^{i\left(\ke(\zj-a)+\km(z-\zj+a)\vphantom{b^2}\right)}.}
Again invoking the fact that the particles are identical, we may remove the index from the forward scattering amplitude and replace the sum over $\ell$ with a factor \smash{$N-1\approx N$}. We thus arrive at the expression
\eq{\Ee_0\,e^{i\left(\ke(\zj-a)+\km(z-\zj+a)\vphantom{b^2}\right)}=\left(\Ei_0+\frac{2\pi i\rho\left(e^{i(\ke-\km)(\zj-3a)}-e^{-i(\ke-\km)a}\right)}{(\ke-\km)\km\.^2}\,\Sfwd\Ee_0\right)e^{i\km z},\label{eq-sc}}
where $\rho=N/V$ is the particle number density. Here we see that our treatment of the problem is indeed consistent to second order in $\rho$, as claimed in section~\ref{sec-model}, for, if we substitute this expression into equation~\ref{eq-Ecint} and replace the sum over $j$ with a factor $N$ and $g\evalrj$ with $1/V$, we can immediately see that our expression for the coherent field (i.e.~the average total field) is quadratic in $\rho$.

Equation~\ref{eq-sc} may be simplified by dividing throughout by $e^{i\km z}$. This yields
\eq{\Ee_0\,e^{i(\ke-\km)(\zj-a)}=\Ei_0+\frac{2\pi i\rho\left(e^{i(\ke-\km)(\zj-3a)}-e^{-i(\ke-\km)a}\right)}{(\ke-\km)\km\.^2}\,\Sfwd\Ee_0,}
which can be split into two even simpler expressions by noting that the equality with only the terms with $e^{i(\ke-\km)\zj}$ and the equality with only the remaining terms must be simultaneously satisfied if the full equality is to be satisfied for every value of $\zj$:
\eq{\Ee_0\,e^{-i(\ke-\km)a}\,e^{i(\ke-\km)\zj}=\frac{2\pi i\rho\,e^{-3i(\ke-\km)a}\,e^{i(\ke-\km)\zj}}{(\ke-\km)\km\.^2}\,\Sfwd\Ee_0\label{eq-sc1}}
and
\eq{0=\Ei_0-\frac{2\pi i\rho\,e^{-2i(\ke-\km)a}}{(\ke-\km)\km\.^2}\,\Sfwd\Ee_0.}
The latter of these is related to the Ewald-Oseen theorem and is of not interest to us in this work. The former yields
\eq{\ke=\km\left(1+\frac{2\pi i\rho\Sfwd\,e^{-2i(\ke-\km)a}}{\km\.^3}\right)\!,\label{eq-master}}
which is enough to solve for the effective propagation constant $\ke$ without necessarily knowing \smash{$\Ee_0$}.

As a first approximation, we may assume that $\ke$ and $\km$ are not too dissimilar and remove the exponential on the right side. This yields van de Hulst's formula for the effective propagation constant\cite{ref-vandeHulst-LSSP}:
\eq{\ke=\km\left(1+\frac{2\pi i\rho\Sfwd}{\km\.^3}\right)\!.\label{eq-vdH}}

As a correction to this, let us iterate once and insert van de Hulst's formula in place of $\ke$ on the right side of equation~\ref{eq-master}, which yields, once we divide by $k$ to obtain the effective refractive index of the suspension,
\eq{\ne=\nm\left(1+\frac{2\pi i\rho\Sfwd\,e^\frac{4\pi\rho\Sfwd a}{\km\.^2}}{\km\.^3}\right).\label{eq-ne}}
This is our principal result.

One might consider the possibility of iterating further and replacing $\ke$ on the right side of equation~\ref{eq-master} with the expression that comes from equation~\ref{eq-ne}. However, as we have stated, our model is self-consistent to second order. By doing this and performing the corresponding algebra, it is easy to see that successive iterations will not affect the term which is of second order in $\rho$ and $\Sfwd$, only higher-order terms. We therefore deem further iterations unnecessary. In fact, for values of \smash{$f=\rho v$} (where $v$ is the volume of one particle) very close to dense packing, it could be detrimental to continue iterating, as it could introduce spurious corrections to high-order terms. We believe that such high densities would require a higher-order treatment of the problem from the beginning (equation~\ref{eq-Egeneral}) and that in this case $N$-particle correlations (with $N$ higher than 2) would become important. It will be seen in section~\ref{sec-exp} that this first iteration is enough to reproduce experimental results accurately.

For non-spherical particles, the forward scattering amplitude of a particle depends on its orientation. For random systems where particle orientation and position are uncorrelated, one must thus average $\Sfwd$ over all the possible particle orientations and use that in place of $\Sfwd$ in equation~\ref{eq-ne}:
\eq{\langle\Sfwd\rangle=\integral{}{}{P(\Omega)\Sfwd}{\Omega},}
where $\Omega$ represents the orientation of a particle and $P(\Omega)$ is the probability density function for particle orientation.

For small values of $\rho\Sfwd$, equation~\ref{eq-ne} reduces to van~de~Hulst's formula. Note that this does not necessarily imply small $\rho$.

This work and our previous work on the topic\cite{ref-NahmadRohen-JOSAA38} have constituted a long, but unambiguous, road, but we are finally at the destination. In the following sections, we will compare our formula for the effective refractive index (equation~\ref{eq-ne}) with experimental results and discuss the formula's validity. In the process, we will provide some insight into the behaviour of the effective refractive index.

\twocolumn

\section{Comparison with experiment}\label{sec-exp}

\begin{figure}[b!]
	\begin{center}
		\begin{tikzpicture}
			\pgfplotsset{every axis legend/.append style={at={(0.5,-0.2)},anchor=north}}
			\begin{axis}[
				width=\graphwidth,
				height=\graphheight,
				legend cell align=left,
				legend style={draw=none},
				xlabel=$f$,
				xmin=0,
				xmax=0.1,
				ylabel=$\Imag{\ne}$,
				ymin=0,
				ymax=0.02,
				scaled ticks=false,
				xticklabel style={/pgf/number format/.cd,fixed,fixed zerofill,precision=2},
				yticklabel style={/pgf/number format/.cd,fixed,fixed zerofill,precision=3},
				axis x line=bottom,
				x label style={at={(ticklabel* cs:1)},anchor=west},
				axis y line=left,
				y label style={at={(current axis.above origin)},anchor=south,rotate=-90}
			]
				\addplot[smooth,thick,blood]file[x index=0,y index=1]{ne-Im-RV-a240.txt};
				\addplot[smooth,thick,sodium]file[x index=0,y index=1]{ne-Im-RV-a550.txt};
				\addplot[smooth,thick,chlorophyll]file[x index=0,y index=1]{ne-Im-RV-a1010.txt};
				\addplot[smooth,thick,blood,dashed]file[x index=0,y index=1]{ne-Im-vdH-a240.txt};
				\addplot[smooth,thick,sodium,dashed]file[x index=0,y index=1]{ne-Im-vdH-a550.txt};
				\addplot[smooth,thick,chlorophyll,dashed]file[x index=0,y index=1]{ne-Im-vdH-a1010.txt};
				\addplot[only marks,mark size=2pt,blood]file[x index=0,y index=1]{ne-Ishimaru-a240.txt};
				\addplot[only marks,mark size=2pt,sodium]file[x index=0,y index=1]{ne-Ishimaru-a550.txt};
				\addplot[only marks,mark size=2pt,chlorophyll]file[x index=0,y index=1]{ne-Ishimaru-a1010.txt};
				\legend{
					$a=240.5$~nm ($ka=2.39$),
					$a=550.5$~nm ($ka=5.46$),
					$a=1\;010$~nm ($ka=10.03$),
				}
			\end{axis}
		\end{tikzpicture}
		\captionsetup{singlelinecheck=off}
	\end{center}
	\caption[.]{$\Imag{\ne}$ as a function of the volume fraction $f$ for monodisperse suspensions of spherical latex particles with different radii $a$ suspended in water and illuminated by light of wavelength $\lambda=633$~nm: comparison of our model (smooth curves), van~de~Hulst's model (dashed curves) and measurements by Ishimaru \& Kuga (filled circles).}
	\label{fig-neIm}
\end{figure}
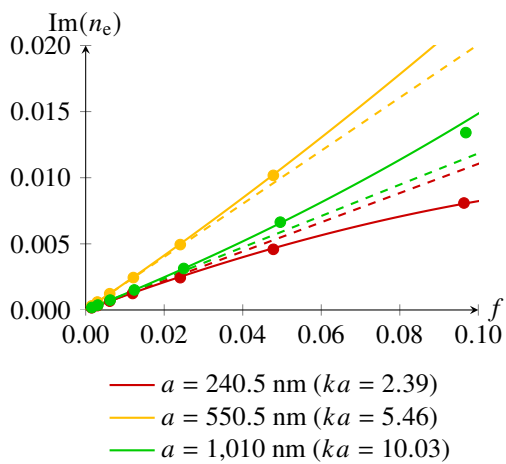

Any model must be compared to experiment in order for its validity (or lack thereof) to be established. There is a surprising scarcity of measurements of the effective refractive index of dense suspensions; to our knowledge, the only such measurements that can be directly compared to our model are those of the imaginary part of the effective refractive index $\ne$ at a wavelength of 633~nm for a suspension of latex particles (particle refractive index \smash{$\np=1.59$}) in water (matrix refractive index \smash{$\nm=1.33$}) taken by A~Ishimaru and Y~Kuga in 1982~\cite{ref-Ishimaru-JOSA72}. Ishimaru and Kuga performed measurements on monodisperse suspensions of spherical particles with radius $a$ between 45.5~nm and about 6~\textmu m. Of the seven particle sizes explored by them, those which are comparable to or larger than the wavelength are appropriate for testing our model, since, for the refractive-index contrast
\eq{\Delta_n=\left|\frac{\np}{\nm}-1\right|}
of their suspensions, scattering is mostly in the forward direction in the \smash{$ka\sim1$} and \smash{$ka\gg1$} regimes (see section~\ref{sec-Sfwd}); lateral scattering and back-scattering are very strong in the \smash{$ka\ll1$} regime. We therefore discard the two smallest particle sizes studied by them. We are unable to make a meaningful comparison of our model with Ishimaru and Kuga's suspensions with the largest particles due to the very large uncertainties in both particle radius and particle refractive index reported by them. Therefore, we also discard the largest particle size.

A more useful quantity than the particle number density $\rho$ in the context of effective-medium models is the (dimensionless) volume fraction $f=\rho v$, where $v$ is the volume of a single particle. For spherical particles, equation~\ref{eq-ne} in terms of the volume fraction is
\eq{\ne=\nm\left(1+\frac{3if\Sfwd\,e^\frac{3f\Sfwd}{(\km a)^2}}{2(\km a)^3}\right).\label{eq-nesph}}
This is the form of $\ne$ we will compare to van de Hulst's formula and to Ishimaru and Kuga's measurements in this section.

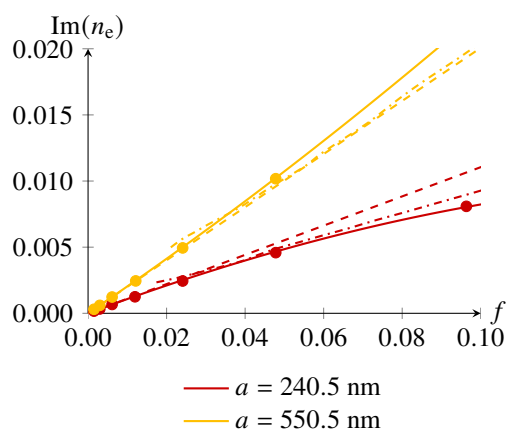
\begin{figure}[b!]
	\begin{center}
		\begin{tikzpicture}
			\pgfplotsset{every axis legend/.append style={at={(0.5,-0.2)},anchor=north}}
			\begin{axis}[
				width=\graphwidth,
				height=\graphheight,
				legend cell align=left,
				legend style={draw=none},
				xlabel=$f$,
				xmin=0,
				xmax=0.1,
				ylabel=$\Imag{\ne}$,
				ymin=0,
				ymax=0.02,
				scaled ticks=false,
				xticklabel style={/pgf/number format/.cd,fixed,fixed zerofill,precision=2},
				yticklabel style={/pgf/number format/.cd,fixed,fixed zerofill,precision=3},
				axis x line=bottom,
				x label style={at={(ticklabel* cs:1)},anchor=west},
				axis y line=left,
				y label style={at={(current axis.above origin)},anchor=south,rotate=-90}
			]
				\addplot[smooth,thick,blood]file[x index=0,y index=1]{ne-Im-RV-a240.txt};
				\addplot[smooth,thick,sodium]file[x index=0,y index=1]{ne-Im-RV-a550.txt};
				\addplot[smooth,thick,blood,dashed]file[x index=0,y index=1]{ne-Im-vdH-a240.txt};
				\addplot[smooth,thick,blood,dashdotted]file[x index=0,y index=1]{ne-simIshimaru-a240.txt};
				\addplot[only marks,mark size=2pt,blood]file[x index=0,y index=1]{ne-Ishimaru-a240.txt};
				\addplot[smooth,thick,sodium,dashed]file[x index=0,y index=1]{ne-Im-vdH-a550.txt};
				\addplot[smooth,thick,sodium,dashdotted]file[x index=0,y index=1]{ne-simIshimaru-a550.txt};
				\addplot[only marks,mark size=2pt,sodium]file[x index=0,y index=1]{ne-Ishimaru-a550.txt};
				\legend{
					{$a=240.5$~nm},
					{$a=550.5$~nm}
				}
			\end{axis}
		\end{tikzpicture}
		\captionsetup{singlelinecheck=off}
	\end{center}
	\caption[.]{$\Imag{\ne}$ as a function of the volume fraction $f$ for monodisperse suspensions of spherical latex particles of size comparable to the wavelength suspended in water and illuminated by light of wavelength $\lambda=633$~nm: comparison of our model (smooth curves), van de Hulst's model (dashed curves), simulations by Ishimaru et al.\ (dash-dotted curves) and measurements by Ishimaru \& Kuga (filled circles).}
	\label{fig-neIm12}
\end{figure}

Figure~\ref{fig-neIm} shows a comparison of theoretical and experimental values of the imaginary part of the effective refractive index $\ne$ of Ishimaru's suspensions with intermediate values of $ka$ (i.e.~particles whose size is comparable to the wavelength). The experimental values of $\Imag{\ne}$ were calculated from the data reported by Ishimaru and Kuga. Because van~de~Hulst's model is linear in $\rho$ (or $f$), it can never take into account dependent scattering. It can be seen that our model correctly predicts the deviation of $\Imag{\ne}$ from linearity both in magnitude and in sign. Importantly, our model does not contain any adjustable parameters, but rather is derived from fundamental scattering theory under the assumption that the particles scatter mostly in the forward direction, so the strength of the agreement between it and experiment is remarkable.

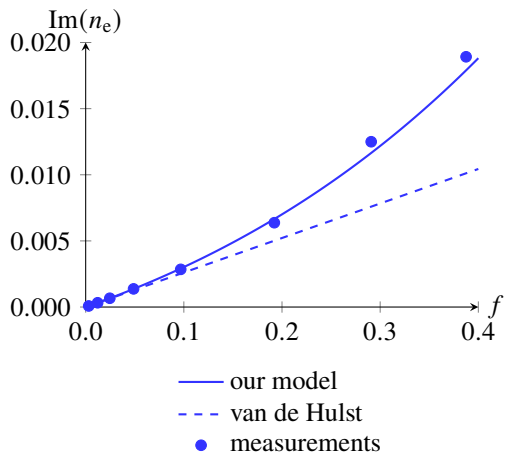
\begin{figure}[b!]
	\begin{center}
		\begin{tikzpicture}
			\pgfplotsset{every axis legend/.append style={at={(0.5,-0.2)},anchor=north}}
			\begin{axis}[
				width=\graphwidth,
				height=\graphheight,
				legend cell align=left,
				legend style={draw=none},
				xlabel=$f$,
				xmin=0,
				xmax=0.4,
				ylabel=$\Imag{\ne}$,
				ymin=0,
				ymax=0.02,
				scaled ticks=false,
				xticklabel style={/pgf/number format/.cd,fixed,fixed zerofill,precision=1},
				yticklabel style={/pgf/number format/.cd,fixed,fixed zerofill,precision=3},
				axis x line=bottom,
				x label style={at={(ticklabel* cs:1)},anchor=west},
				axis y line=left,
				y label style={at={(current axis.above origin)},anchor=south,rotate=-90}
			]
				\addplot[smooth,thick,cobalt]file[x index=0,y index=1]{ne-Im-RV-a2850.txt};
				\addplot[smooth,thick,cobalt,dashed]file[x index=0,y index=1]{ne-Im-vdH-a2850.txt};
				\addplot[only marks,mark size=2pt,cobalt]file[x index=0,y index=1]{ne-Ishimaru-a2850.txt};
				\legend{
					our model,
					van~de~Hulst,
					measurements
				}
			\end{axis}
		\end{tikzpicture}
		\captionsetup{singlelinecheck=off}
	\end{center}
	\caption[.]{$\Imag{\ne}$ as a function of the volume fraction $f$ for monodisperse suspensions of spherical latex particles of radius $a=2\;850$~nm suspended in water and illuminated by light of wavelength $\lambda=633$~nm: comparison of our model, van~de~Hulst's formula and measurements by Ishimaru \& Kuga over an extended $f$ range.}
	\label{fig-neIm4}
\end{figure}

In 1983, Ishimaru and colleagues presented the results of a computer simulation meant to validate the experimental results for the four lowest of the seven particle sizes Ishimaru and Kuga had studied the previous year~\cite{ref-Varadan-RS18}. Figure~\ref{fig-neIm12} shows a comparison of our model to Ishimaru et al's simulation for the two smallest particle sizes we are considering here (which are the two largest particle sizes they ran the simulation for). While the simulation certainly approaches the experimental data, our model does so more closely.

Our model also agrees remarkably well with Ishimaru and Kuga's measurements of suspensions of particles with a radius of $2.85$~\textmu m (figure~\ref{fig-neIm4}). We are fortunate in that here Ishimaru and Kuga performed measurements up to $f\approx0.38$, a value much higher than effective-medium models are usually considered valid for. There is very good agreement between theory and experiment even for these high values of $f$, with the model predicting a value of $\Imag{\ne}$ only 5.4\% lower than measured for the highest value of $f$ experimental data exists for (and this assuming the measurements were more precise than $10^{-5}$ in the imaginary part of the refractive index). This particle size is especially interesting given its proximity to the average size of red blood cells ---which occur in the bloodstream at volume fractions of 0.40--0.45 under normal circumstances\cite{ref-Guyton-TMP}--- and the droplets in topical creams --- which can reach volume fractions of 0.5\cite{ref-Tadros-E}. In both of these cases, the refractive-index contrast between particles and matrix is even lower than in the case of latex particles suspended in water (the effect of particle size and refractive-index contrast on how much electromagnetic energy is scattered in the forward direction is discussed in detail in section~\ref{sec-Sfwd}).

While the two-particle correlation function is present in our derivation, we have argued on physical grounds that its precise form becomes increasingly unimportant as the particles in the suspension become increasingly forward-scattering. The close agreement of our formula with experimental data is an indicator that this is indeed the case.


\section{The meaning of forward scattering and further discussion}\label{sec-Sfwd}

\begin{figure}[t!]
	\begin{center}
		\begin{tikzpicture}
			\pgfplotsset{every axis legend/.append style={at={(-0.075,0.5)},anchor=east}}
			\begin{polaraxis}[
				width=\graphwidth,
				height=\graphwidth,
				legend cell align=left,
				legend style={draw=none},
				xmin=0,
				xmax=360,
				xticklabels={,
					0,
					$\displaystyle\frac{\pi}{6}$,
					$\displaystyle\frac{\pi}{3}$,
					$\displaystyle\frac{\pi}{2}$,
					$\displaystyle\frac{2\pi}{3}$,
					$\displaystyle\frac{5\pi}{6}$,
					$\pi$,
					$\displaystyle\frac{5\pi}{6}$,
					$\displaystyle\frac{2\pi}{3}$,
					$\displaystyle\frac{\pi}{2}$,
					$\displaystyle\frac{\pi}{3}$,
					$\displaystyle\frac{\pi}{6}$
				},
				ymin=0,
				ymax=1,
				y axis line style={opacity=0},
				yticklabels=,
				scaled ticks=false,
				major tick length=0
			]
				\addplot[smooth,thick,iodine]file[x index=0,y index=1]{S-Mie-a54.txt};
				\addplot[smooth,thick,blood]file[x index=0,y index=1]{S-Mie-a240.txt};
				\addplot[smooth,thick,sodium]file[x index=0,y index=1]{S-Mie-a550.txt};
				\addplot[smooth,thick,chlorophyll]file[x index=0,y index=1]{S-Mie-a1010.txt};
				\addplot[smooth,thick,cobalt]file[x index=0,y index=1]{S-Mie-a2850.txt};
				\legend{
					$ka=0.54$,
					$ka=2.39$,
					$ka=5.46$,
					$ka=10.03$,
					$ka=28.29$,
				}
			\end{polaraxis}
		\end{tikzpicture}
		\begin{tikzpicture}
			\pgfplotsset{every axis legend/.append style={at={(0.024,0.5)},anchor=east}}
			\begin{polaraxis}[
				width=\graphwidth,
				height=\graphwidth,
				legend cell align=left,
				legend style={draw=none},
				xmin=0,
				xmax=360,
				xticklabels={,
					0,
					$\displaystyle\frac{\pi}{6}$,
					$\displaystyle\frac{\pi}{3}$,
					$\displaystyle\frac{\pi}{2}$,
					$\displaystyle\frac{2\pi}{3}$,
					$\displaystyle\frac{5\pi}{6}$,
					$\pi$,
					$\displaystyle\frac{5\pi}{6}$,
					$\displaystyle\frac{2\pi}{3}$,
					$\displaystyle\frac{\pi}{2}$,
					$\displaystyle\frac{\pi}{3}$,
					$\displaystyle\frac{\pi}{6}$
				},
				ymin=0,
				ymax=1,
				y axis line style={opacity=0},
				yticklabels=,
				scaled ticks=false,
				major tick length=0
			]
				\addplot[smooth,thick,blood]file[x index=0,y index=1]{S-Mie-a240.txt};
				\addplot[blood,draw opacity=0,opacity=0.3,fill]file[x index=0,y index=1]{S-Mie-a240-pi6.txt}\closedcycle;
			\end{polaraxis}
		\end{tikzpicture}
		\captionsetup{singlelinecheck=off}
	\end{center}
	\caption[.]{Top: $|\Stheta|^2/|\Sfwd|^2$ for different values of $ka$ as calculated using Mie's equations. The values of $ka$ correspond to five of the particle sizes used in Ishimaru's study for a wavelength of 633~nm and a refractive-index contrast \smash{$\Delta_n=0.2$}\cite{ref-Ishimaru-JOSA72}; the smallest value is shown as an example of a case in which the particles are not forward-scattering and our model is not applicable, while the other four values correspond to the suspensions analysed in figures~\ref{fig-neIm}--\ref{fig-neIm4}. The radial scale ranges from 0 to 1. Bottom: As top for the specific case $ka=2.39$. The region $\theta\leq\pi/6$ is shaded.}
	\label{fig-Mie}
\end{figure}
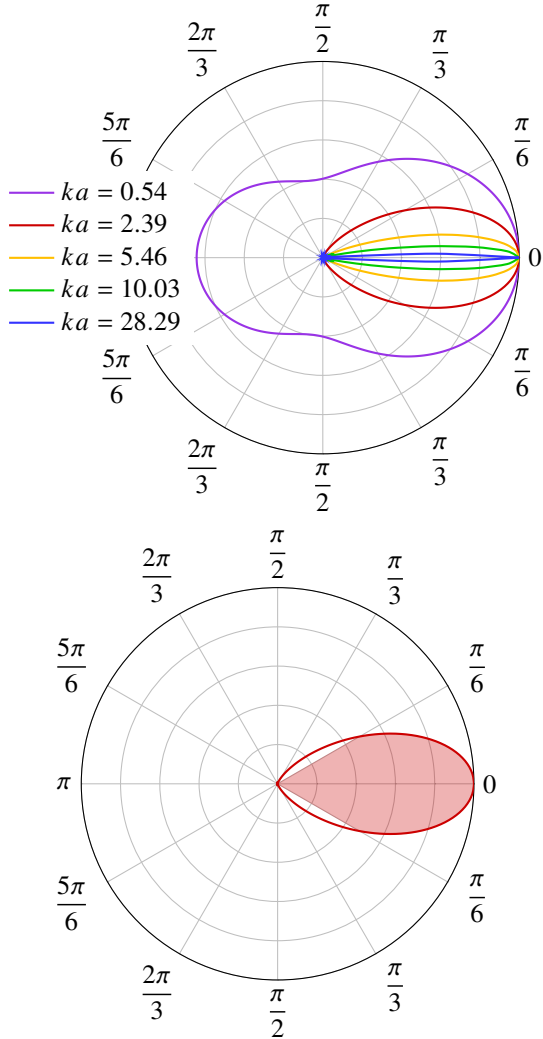

Certain types of suspensions of practical interest consist of particle-matrix combinations which result in the scattering being mostly in the forward direction. Some examples are the suspensions used in Ishimaru and Kuga's work, some emulsions\cite{ref-Tadros-E} and many types of biological tissue\cite{ref-Xu-OL30}. Contrary to popular belief, the list of suspensions in which the particles scatter mostly in the forward direction is not restricted to suspensions of large tenuous particles; for example, for a latex particle ($\np=1.59$) with a radius of 38\% of the wavelength (i.e.~the particle is somewhat smaller than the wavelength) suspended in water ($\nm=1.33$), where the refractive-index contrast is \smash{$\Delta_n\approx0.2\not\ll1$} (i.e.~the particle is not tenuous), 77.5\% of the scattered energy is scattered at angles smaller than $\pi/6$ and 93.0\% of it is scattered at angles smaller than $\pi/4$ (see figure~\ref{fig-Mie}). Even for suspensions of such non-large, non-tenuous particles, our model shows excellent agreement with experiment, as shown in section~\ref{sec-exp} (figures~\ref{fig-neIm} and~\ref{fig-neIm12}).

\begin{figure}[t!]
	\begin{center}
		\begin{tikzpicture}
			\begin{axis}[
				width=\graphwidth,
				height=\graphwidth,
				axis line style=thick,
				legend cell align=left,
				legend style={draw=none},
				xlabel=$\Delta_n$,
				x label style={at={(current axis.right of origin)},anchor=west},
				ylabel={$ka$},
				y label style={at={(current axis.above origin)},anchor=south,rotate=-90},
				enlargelimits=false,
				colorbar
			]
				\addplot graphics[
				xmin=0,
				xmax=2,
				ymin=0,
				ymax=35
				]{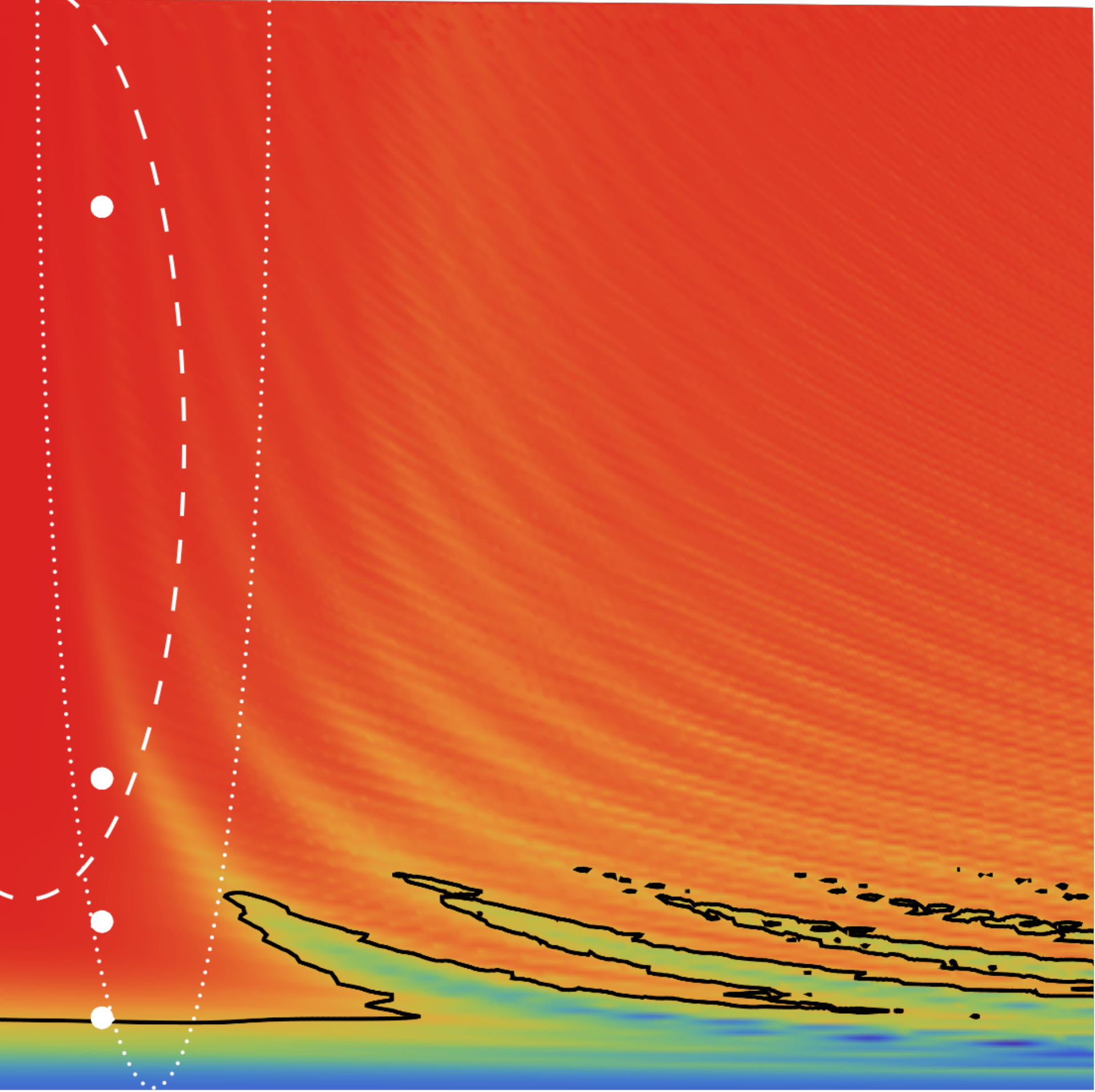};
			\end{axis}
		\end{tikzpicture}
		\captionsetup{singlelinecheck=off}
	\end{center}
	\caption[.]{Fraction of the total scattered energy which is scattered approximately in the forward direction (at angles smaller than $\pi/6$) as a function of the refractive-index contrast $\Delta_n$ and the particle size parameter $ka$. The black contour separates the region in which this fraction is more than 75\% and thus our model is valid (top) from the region where this fraction is less than 75\% and thus our model is likely not applicable. Filled circles indicate the refractive-index contrast and size parameter for Ishimaru and Kuga's suspensions studied in this work; the white dashed contour indicates the region where these values fall for biological tissue, while the white dotted contour indicates where they fall for certain emulsions (see main text).}
	\label{fig-goodness}
\end{figure}

Figure~\ref{fig-goodness} shows the fraction of the total scattered energy which is scattered at angles smaller than $\pi/6$ for different values of the refractive-index contrast $\Delta_n$ and the particle size parameter $ka$. The oscillations in the amount of forward-scattered energy are a consequence of resonances in the Mie coefficients due to the spherical Bessel and Hankel functions involved. The black contour in the graph is located at a value of 75\%. The particles of any suspension which falls on the region above this contour can be considered to be forward-scattering particle, and thus all such suspensions are amenable to our model. Also shown in the graph are the values of $\Delta_n$ and $ka$ for some of the suspensions studied by Ishimaru and Kuga which we have touched on here (the two suspensions studied by them with the smallest particle radii fall below the contour and cannot be modelled with our present formula), as well as the region in which the values for biological tissue\cite{ref-Xu-OL30,ref-Lazareva-JBO23,ref-Borovoi-PSPIE3194,ref-Myakov-JBO7,ref-Brunsting-BJ14,ref-Zhang-SR7,ref-Malik-JHC46,ref-Wilson-AIB} and some emulsions (such as those used for parenteral nutrition, blood substitutes and topical creams)\cite{ref-Tadros-E,ref-Sano-JCIS124,ref-Li-S19,ref-Rheims-MST8} fall. Most of these points fall deep within the forward-scattering region of $(\Delta_n,ka)$ space (that is, the region of validity of our model), with only one of Ishimaru and Kuga's suspensions being somewhat close to the region's boundary but still within the region.

\section{Conclusions}\label{sec-conc}

We have derived a model for the effective refractive index $\ne$ of a suspension of identical particles. The model was derived unambiguously from fundamental scattering theory assuming only that the particles in question scatter mostly in the forward direction; it has the advantage of providing an analytical expression for $\ne$ rather than being a conceptually difficult and computationally costly numerical simulation. It is self-consistent up to terms of second order in the particle number density $\rho$ (or, equivalently, in the volume fraction $f$). Though the formula is the same one that appeared in a previous work\cite{ref-NahmadRohen-JOSAA38}, the result is much more general, as it does not require that the particles be large and tenuous, merely that they scatter mostly in the forward direction, and the derivation presented here is much more transparent both mathematically and physically. For instance, although the two-particle correlation function is present in our derivation, we have shown that the forward-scattering assumption makes the form this function takes unimportant. Our model's consistency with experimental data confirms this.

We have shown that our model reproduces remarkably well the (admittedly scarce) experimental data on $\Imag{\ne}$ for suspensions of medium and large forward-scattering particles available in the scientific literature; this is true even for dense suspensions where almost 40\% of the volume is taken up by the particles. Whether the model can also accurately calculate $\Real{\ne}$ and whether a similar model can properly calculate the effective refractive index of semi-ordered and ordered suspensions are things that remain to be seen; this will require appropriate experimental data to become available. Presently, however, this proves that $\Imag{\ne}$ can be used as a tool for the diagnosis of dense suspensions of forward-scattering particles.

Because our model is applicable to any suspension of forward-scattering particles, it is important to know exactly what ``forward-scattering particles'' means in this context. Previous works\cite{ref-NahmadRohen-PS91,ref-NahmadRohen-JOSAA38,ref-Twersky-JOSA60,ref-Twersky-JOSA52,ref-Twersky-JMP3} have restricted this to particles which are simultaneously large with respect to the wavelength ($ka\gg1$) and tenuous ($\Delta_n\ll1$), for which the anomalous-diffraction approximation is valid (here, however, we do not require this approximation). Our comparison with experiment and subsequent analysis of the angular scattering of light by the types of particles involved in said experiment have shown that a less restrictive, but still appropriate, definition is ``any particle for which at least 75\% of its scattered light is scattered at angles below $\pi/6$''. We have shown a map of a region in $(\Delta_n,ka)$ space which is of interest in various fields of science, engineering and medicine and marked not only which portion of this region corresponds to forward-scattering particles, but also where certain highly important types of suspension fall on the map. The model we have presented is therefore applicable to a wide range of suspensions, including those with high volume fractions, such as artificial materials of particular industrial, chemical and pharmaceutical interest\cite{ref-tadros-E} to most kinds of biological tissue\cite{ref-Xu-OL30}.

There is no reason our model should not be applicable to suspensions with absorbing matrix and/or particles; nowhere in our derivation have we assumed $\nm$ and $\np$ are real. Determining the region of validity of the model, however, requires experimental data on such suspensions to become available.

We strongly believe this work can contribute to all-optical, non-destructive, rapid diagnosis of some blood diseases, determination of food quality and structural analysis of artificial materials, to name just a few applications.

\section{Acknowledgements}

The authors would like to thank Anays Acevedo-Barrera, Rub\'en Barrera-P\'erez and Omar Wilfrido V\'azquez Estrada for interesting and valuable discussions which considerably improved the quality of the present work.

The authors also acknowledge financial support from Direcci\'on General de Asuntos del Personal Acad\'emico from Universidad Nacional Aut\'onoma de M\'exico through project IN101821 and through a postdoctoral contract for ANR.

\section{Disclosures}

The authors declare no conflicts of interest.

\section{References}

\begingroup
\renewcommand{\section}[2]{}

\endgroup

\vfill

\end{document}